\documentstyle[aps,epsf]{revtex}
\gdef\journal#1, #2, #3, 1#4#5#6{		% Journal reference.  Comma sets
    {\sl #1~}{\bf #2}, #3 (1#4#5#6)}		% off: name, vol, page, year
\def\commpureapplmath{\journal Commun. Pure Appl. Math., }
\def\cpl{\journal Chem. Phys. Lett., }

\begin{document}
\title{
%{\it Rough draft, to be submitted to JCP, Feb. 1997}
%\vskip 2mm
Accuracy of Electronic Wave Functions in Quantum Monte Carlo:
the Effect of High-Order Correlations}
\author{Chien-Jung Huang}
\address{Laboratory of Atomic and Solid State Physics,\\
Cornell University, Ithaca, NY 14853}
\author{C.J. Umrigar}
\address{Cornell Theory Center and Laboratory of Atomic and Solid State Physics,\\
Cornell University, Ithaca, NY 14853}
\author{M.P. Nightingale}
\address{Physics Department, University of Rhode Island,\\ Kingston,
Rhode Island 02881}

\maketitle
\begin{abstract}

Compact and accurate wave functions can be constructed by quantum
Monte Carlo methods.  Typically, these wave functions consist of a sum
of a small number of Slater determinants multiplied by a Jastrow
factor.  In this paper we study
the importance of including high-order, nucleus-three-electron
correlations in the Jastrow factor.
An efficient algorithm based on the theory of invariants is used to
compute the high-body correlations.
We observe significant
improvements in the variational Monte Carlo energy and in the
fluctuations of the local energies but not in
the fixed-node diffusion Monte Carlo energies.
Improvements for the ground states
of physical, fermionic atoms are found to be smaller than those for
the ground states of fictitious, bosonic atoms, indicating that
errors in the nodal surfaces of the fermionic wave functions are
a limiting factor.
\end{abstract}

\section{Introduction}
\label{Intro}

Optimized trial wave functions that closely approximate eigenstates of
Hamiltonians are essential ingredients of accurate electronic
structure calculations employing quantum Monte Carlo methods.  The quality of
these trial wave functions is relevant both for expectation values of
physical interest and for the variance of the Monte Carlo estimators,
which is a measure of the efficiency of the computation.

When one uses a variational Monte Carlo method, as the trial wave
function approaches an exact eigenstate of the Hamiltonian, the energy
and expectation values of quantities commuting with the Hamiltonian
satisfy a zero-variance principle, i.e., the expectation values
approach the exact eigenstate values, while the Monte Carlo variance
goes to zero.  More sophisticated forms of quantum Monte Carlo, such
as diffusion Monte Carlo, attempt to project out the ground state from
the trial state, in which case exact expectation values are obtained
for observables that commute with the Hamiltonian, even if the trial
state is not an exact eigenstate. In this case, the quality of the
trial states affects only the statistical errors.  However, most
practical algorithms suppress admixtures of excited states completely
only for nodeless wave functions, such as bosonic ground states.  For
trial functions with nodes, usually the fixed-node
approximation\cite{fixednode} is made, in which case errors in the
nodal surface systematically bias the expectation values.  The usual
variational and mixed-estimators\cite{mixedestim} of observables that
do not commute with the Hamiltonian always yield results for which the
magnitude of the bias of the expectation values and also the
statistical errors depend on the quality of the trial wave function but
no zero-variance principle is satisfied in this case.  In sum, for the
commonly used forms of quantum Monte Carlo, it is essential to employ
accurate trial wave functions, in particular in the case of operators
that do not commute with the Hamiltonian.

The wave functions used in electronic structure calculations employing
quantum Monte Carlo usually consist of a product of a Jastrow factor
and one or a sum of Slater determinants.  The simplest, and possibly
most commonly used wave function of this sort consists of a single
determinant multiplied by a simple Jastrow factor that is a product
over electron pair contributions.  As regards the physics contained in
the determinantal and Jastrow factors, it is generally believed that
multiple determinants most efficiently incorporate near-degeneracy or
non-dynamic correlation, while a Jastrow factor efficiently supplies
the major portion of the dynamic correlation.  It has been
shown\cite{UWW} that a Jastrow factor that correlates two electrons
and a nucleus gives much better variational energies and has much
smaller fluctuations of the local energy than a Jastrow factor
correlating only pairs of electrons.  Given this success, one may well
ask whether it is advantageous to include the next most important
correlations, {\it viz.} four-body correlations of three electrons and
a nucleus.

There are different measures of success that one can use to answer
this question.  The criteria we shall use in this paper are the
reduction in the variational energy and in the fluctuations in the
local energy.  It is reasonable to equate a reduction in these two
quantities to improvement of the quality of the wave function and to a
reduction of the systematic error in expectation values of operators
that do not commute with the Hamiltonian.  We note that, as far as
diffusion Monte Carlo is concerned, a reduction of the fluctuations in
the local energy also has the important advantage that the time-step 
error is usually reduced also\cite{ourDMC}.

An alternative measure of the success of modification of the trial
wave function is the improvement in the fixed-node estimate of the
energy.  The wave functions used in this paper were obtained by
simultaneous optimization of parameters appearing in the determinantal
and Jastrow factors.  For fixed parameters, the former determines the
location of the nodal surface and the value of the fixed-node energy.
The optimization feeds back changes in the Jastrow factor to the
location of the nodal surface.  We observe in the work
reported in this paper that this has only a small effect on
the fixed-node energy.  Apparently, the nodal surface changes
little even for modifications of the Jastrow factor that greatly
improve the wave function.

The purpose of this paper is to study the effect of incorporation of
many-body correlations in the Jastrow factor.  In Sec.~\ref{WaveFunc}
we present the form of the trial wave functions and discuss the cusp
conditions imposed on the wave functions, which reduce greatly the
number of free parameters to be optimized.  To facilitate the
computations, we use bases of invariants, similar to those introduced by Mushinski
and Nightingale\cite{Peter_inv} in their study of bosonic van der
Waals clusters.  The basis invariants employed are given in Appendix
\ref{inv}, where we also present an algorithm for using
the invariants to calculate the wave function and its first two
spatial derivatives.  In Sec.~\ref{Result} we present the results for
the Li, Be and Ne atoms, and also for fictitious bosonic Li, Be and Ne
atoms.  These latter model systems are introduced to allow us to
disentangle flaws of the wave function associated with many-body
correlations and the nodal surface.

\section{Functional form of the wave function}
\label{WaveFunc}

The wave functions we use have the form
\begin{eqnarray}
\Psi=
\sum_l d_l D^\uparrow_l D^\downarrow_l \; \prod_n J_n
\label{eq1}
\end{eqnarray}
$D^\uparrow_l$ and $D^\downarrow_l$ are the Slater determinants of
single particle orbitals for the up and down electrons respectively,
and $J_n$ is a Jastrow factor correlating $n$-tuples of electrons and
a nucleus.  The simplest, and possibly most commonly used wave function
of this type contains just a single product of up- and down-spin
determinants and a Jastrow factor that correlates only pairs of
electrons, i.e.,
\begin{eqnarray}
J_2=\prod_{\alpha,i}
\exp A(r_{\alpha i})\prod_{i<j}\;\exp B_{s_{zi}+s_{zj}}(r_{ij}) .
\label{jsim}
\end{eqnarray}
Here, the index $\alpha$ labels the nuclei while $i$ and $j$ label the
electrons; $s_{zk}=\pm{1 \over 2}$ denotes the $z$-component of the
spin of electron $k$, so that the index $t$ of $B_t$ assumes three
values, $t=0,\pm 1$, which in principle allows the correlations
between electron pairs to depend on the orientation of the electron
spins relative to each other and to the fixed $z$-component of the
total electron spin.  Dependence of the function $B_t$ on $t$ allows
for a better variational wave function, as judged by the variational
energy and the fluctuations in the local energy. More specifically, it
is not possible to satisfy both the cusp conditions\cite{Kato} for
parallel and anti-parallel spins unless $B_0 \ne B_{\pm 1}$.
Unfortunately, however, dependence of $B_t$ on $t$ causes {\it spin
contamination}\cite{spin-contam}, i.e., the resulting wave function is
no longer an eigenfunction of the square of the total spin.

Our compromise is to allow for the minimal amount of spin dependence
of the Jastrow factor required to satisfy the cusp conditions, a
parsimonious approach with the added benefit of reducing the number of
variational parameters.
In fact, tests on wave functions with the
additional freedom showed only a small improvement for wave functions
with three-body Jastrow functions and no improvement for wave functions
with four-body Jastrow functions.
To obtain the
results reported here, the electron-electron part of the two-body
Jastrow factor $J_2$ was chosen to contain spin-dependent
coefficients, but not any of the higher-order factors $J_n,\ n>2$.
We use
\begin{eqnarray}
B_t(r_{ij}) = {b_t R_{ij} \over 1 + b'_t R_{ij}},
\label{eq.pairB}
\end{eqnarray}
where $b_0=1/2$ (anti-parallel spins) and $b_{\pm 1}=1/4$ (parallel
spins); further, the
\begin{eqnarray}
R_{ij}\equiv R(r_{ij}) =  (1-e^{-\kappa r_{i j}})/\kappa
\label{eq.scaleR}
\end{eqnarray}
are inter-particle distances scaled by the function $R$, as given on
the right hand side of Eq.~(\ref{eq.scaleR}).  The function $A$ in
Eq.~(\ref{jsim}), which has the same functional form as $B$, but no
spin dependence, and the determinantal part of the wave function are
adjusted to satisfy the electron-nucleus (e-n) cusp condition.

Both the expectation value of the variational energy and the
fluctuations of the local energy can be reduced significantly by going
beyond this simplest form, Eq.~(\ref{jsim}), and it has been
shown\cite{UWW} that a large improvement can be obtained by
generalizing the two-body electron-nucleus Jastrow factor to a
three-body electron-electron-nucleus (e$^2$-n) Jastrow factor
\begin{eqnarray}
J_3=\prod_{\alpha,i<j}\;\exp C(r_{ij},r_{\alpha i},r_{\alpha j}),
\end{eqnarray}
where again the Greek index labels the nuclei and the roman indices
label the electrons.
In the present paper, the function $C$ consists of a 5th order
polynomial in the
scaled inter-particle distances and terms motivated by the Fock
expansion\cite{Fock}.  These terms improve the boundary conditions
satisfied by the trial wave function in reducing the dependence of the
local energy on the shape of the infinitesimal triangle formed by two
electrons and a nucleus, in the limit that two electrons coincide with
a nucleus\cite{MUSM}.  The detailed form of these terms is presented
in Ref.~\onlinecite{Claudia_mol}.

The next step is to introduce a four-body, three-electron-nucleus
(e$^{3}$-n) Jastrow factor
\begin{eqnarray}
J_4=\prod_\alpha \prod_{i<j<k}\;
\exp D(r_{ij},r_{jk},r_{ki},r_{\alpha i},r_{\alpha j},r_{\alpha k}).
\end{eqnarray}
Including these four-body terms had not been done previously in
electronic structure quantum Monte Carlo calculations, but up to
five-body correlations were included in Mushinski and Nightingale's
study of bosonic van der Waals clusters. Of course, inclusion of
higher-body correlations in the Jastrow factor is computationally
expensive, and in Ref.~\onlinecite{Peter_inv} the theory of invariants
was employed to reduce the computational effort.  In principle, for
polynomials of high order, this approach allows one to obtain a
speed-up by a factor equal to the number of elements in the particle
permutation symmetry group associated with the $r_{ij}$ and $r_{\alpha
i}$ in the highest-order correlations included in the Jastrow factor.
In practice, the speed-up is considerably smaller and depends on many
variable details of the computation.  For further more details we
refer to Appendix~\ref{app.combinatorics} and Ref.\cite{Speedup}.

As mentioned above, near-degeneracy correlations such as occur in the
case of Be discussed below can be accounted for
efficiently\cite{UWW,Claudia_mol,HarrisonHandy85,other_multidet} by
inclusion of additional determinants of low-lying orbitals in
Eq.~(\ref{eq1}). In this work, we use four determinants (two configuration
state functions) for Be and one determinant for Li and Ne.

\section{Results and discussion}
\label{Result}
\subsection{Ground state of Li, Be, and Ne}
\label{ground}

In Table~\ref{fermionic}, we show the variational Monte
Carlo energies and the standard deviation of the local energies of Li, Be, and
Ne atoms for wave functions containing four-body correlations and
compare them with the results for two-body and three-body correlated wave functions used
in our earlier work.
As is well-known\cite{UWW} by now, inclusion of the three-body correlations
results in a large improvement in the energy and a large reduction in the
standard deviation of the local energy.  
%For Li, Be and Ne, 86\%, 87\% and 84\%
For Li, Be and Ne, 95\%, 89\% and 84\%
of the correlation energy, missing in the two-body wave functions, is recovered
while the standard deviation of the local energy is reduced by factors of 6, 4 and 2 respectively.
Although not shown in the table, a further advantage accrues from the
fact that the auto-correlation time of the local energies sampled
in variational Monte Carlo\cite{my_accel} and in diffusion Monte
Carlo are somewhat reduced.
In the present work, we observe from Table~\ref{fermionic} that inclusion of
the four-body correlations also results in a significant but smaller improvement
in both the variational energies and the fluctuations of the local energy.
The improvement gets smaller with increasing atomic number and is disappointingly
small for Ne.

In Table~\ref{fermionic}, we also show the diffusion Monte
Carlo energies.  Since the nodes of the wave functions are
determined by the determinantal part of the wave functions only, it is
clear that if the same determinants are used in two wave
functions, the fixed-node diffusion Monte Carlo energies must be the
same to within statistical error.  However, we re-optimized also the
determinantal part of the wave functions when we added the higher-body
terms in the Jastrow exponents, since we had hoped that the additional
freedom would allow the determinantal part of the wave function to
attain a more optimal nodal structure.  However, we find only a small
improvement in the diffusion Monte Carlo energy upon going from the
two-body to the three-body Jastrow, while the difference
of the three-body and four-body cases is
within the statistical error.
The reader should not conclude from these results that, in diffusion
Monte Carlo, improvements in the wave function resulting from improved
Jastrow factors are worthless.  First of all, improved Jastrow factors are likely
to yield more reliable expectation values of operators that do not
commute with the Hamiltonian.  Secondly, even if one is interested only in
the diffusion Monte Carlo energy, the improved Jastrow factor results
in smaller statistical errors for a given number of Monte Carlo steps
and usually also in smaller time-step errors,
leading to a more reliable extrapolation to the zero time-step limit.
We typically find that the time-step error is much smaller for the
three-body Jastrow wave function than for the corresponding two-body Jastrow wave function.
For example, the three-body Jastrow wave function for Be has
a time-step error that is approximately 30 times smaller than
that of the two-body Jastrow wave function. However, the time-step error is of comparable 
magnitude for the three-body and four-body Jastrow wave functions.
The best possible Jastrow factor (one that includes all-body
correlations and has an infinite-order polynomial) would have a
variational Monte Carlo energy that equals the fixed-node diffusion
Monte Carlo energy. Hence, we measure the efficiency of the
four-body contributions to the Jastrow factor by
\begin{eqnarray}
\eta = {E_{\rm VMC}^{\rm 3-body}-E_{\rm VMC}^{\rm 4-body}
\over E_{\rm VMC}^{\rm 3-body}-E_{\rm DMC}}.
\label{efficiency}
\end{eqnarray}

The percent efficiency
rapidly decreases from 60\% for Li to 31\% for Be to 9\% for Ne;
similarly, the reduction in the root mean square fluctuations of the local energy
decreases from 24\% for Li to 12\% for Be to 2\% for Ne.  We attribute
these somewhat disappointing results to flaws of the nodal surface of
the trial wave function,
%%%  which the four-body interactions are not
 which the four-body interaction Jastrow factor is not
designed to correct.  That is, the approximate, fixed-node, diffusion
Monte Carlo wave function has discontinuous derivatives almost
everywhere across the nodal surface.  On the other hand, a Jastrow
factor expressed in terms of inter-particle coordinates, will have
non-analyticities only at the $(3N-3)$-dimensional surface where
particles coincide, which constitutes only a vanishingly small
fraction of the entire $(3N-1)$-dimensional nodal surface.  Hence,
over most of the nodal surface, we are attempting to describe a
non-analytic function as a finite sum of analytic functions and we expect the
convergence to be slow.

To test the validity of the above argument, we also performed
calculations for the nodeless, bosonic ground states of the
Hamiltonians of the same atoms.  From Table~\ref{bosonic} we see that the
efficiency there is considerably greater, 77\% for Li, 60\% for Be
and 50\% for Ne. Also,
the improvement in the fluctuations of the local energy is
considerably larger than in the fermionic case.
% 39,40,33 %.

This supports our conjectured explanation in terms of the nodal
surface.  However, we do find it considerably easier to optimize the
bosonic than the fermionic wave functions.  As a result, we cannot
completely rule out the possibility that the fermionic wave functions
have been optimized to local minima and that considerably better local
minima exist, although we consider it unlikely.

The fact that for bosonic Ne, the e${^3}$-n
correlations already account for 50\% of the missing energy in the e$^2$-n
wave function is in support of our expectation that the low-body order
correlations are the most important ones.  In fact, they may actually
account for more than 50\% if we go beyond polynomial order 5,
since for bosonic Li, where all body-order
correlations are included, the e$^3$-n correlations account only for 77\%
rather than 100\% of the missing energy in the e$^2$-n wave function,
indicating that a higher order polynomial is needed to capture most of the
remaining 23\%.

For any given wave function it is relevant to ask whether further
improvements can be made most economically by increasing the
variational freedom of the determinantal part of the wave function
(either by increasing the number of single-particle basis functions or
by increasing the number of determinants) or by improving the Jastrow
part of the wave function (either by increasing the polynomial-order or
the body-order).  It is clear that if we follow the former route to the
limit of a complete basis of Slater determinants then the exact result
can be obtained, even without a Jastrow factor.  However, the rate of
convergence would be exceedingly slow, because the Slater determinants
lack singularities present in the wave function, such as the cusps at
electron-electron coincidence points.  These cusps can already be built
into the wave function at the level of the two-body correlations by
including a Jastrow factor.
Incorporating three-body correlations is clearly very advantageous,
but the results of this paper show that inclusion of four-body
correlations may not be the most economical next step to further
improvement of the wave functions, at least for the heavier systems.
Instead, it may be preferable to include more determinants in the
wave function.  Once a sufficiently large number of determinants have
been included, it seems likely that it may again become more economical
to improve the Jastrow part by including the four-body correlations.
Such explorations are needed in order to fully exploit
the flexibility that quantum Monte Carlo offers over
conventional quantum chemistry methods for the construction of
accurate, yet relatively compact, wave functions.

\begin{table}[tbh]
\caption[]
{ Total energies of Li, Be, and Ne atoms obtained in variational and
diffusion Monte Carlo. For each atom, the first, second, and third rows
respectively contain results using wave functions with $n$-body
correlations respectively with $n=2$, $n=3$, and $n=4$.
The number of configuration state functions in the determinantal part
of the wave function is denoted by CSF.
$E_0$ is the ``exact'' total energy from Ref.\cite{Exact}. $E_c$ is the
correlation energy.  $E^{\rm VMC}_{c}$ and $E^{\rm DMC}_{c}$ are the
percentages of correlation energy recovered in variational and diffusion Monte
Carlo. $\sigma_{\rm VMC}$ is the root mean square fluctuation of the
local energy in variational Monte Carlo.  The numbers in parentheses
are the statistical errors in the last digit.  The last column is the
efficiency of the four-body correlations as measured in
Eq.(~\ref{efficiency}).  Energies are in Hartree atomic units.  }
\begin{tabular}{cccdddddddc}
  \multicolumn{1}{c} {\ Atom}  &
  \multicolumn{1}{c} {$n$}     &
  \multicolumn{1}{c} {CSF}     &
  \multicolumn{1}{c} {\hskip0.8cm $E_0$}   &
  \multicolumn{1}{c} {\hskip.6cm $E_{\rm c}$} &
  \multicolumn{1}{c} {\hskip.9cm $E^{\rm VMC}$} &
  \multicolumn{1}{c} {\hskip.1cm $E_c^{\rm VMC}$ (\%)} &
  \multicolumn{1}{c} {\hskip.9cm $E^{\rm DMC}$} &
  \multicolumn{1}{c} {\hskip.1cm $E_c^{\rm DMC}$ (\%)} &
  \multicolumn{1}{c} {\hskip.4cm $\sigma_{\rm VMC}$} &
  \multicolumn{1}{c} {$\eta$} \\[.05cm]
\hline
\\[-.2cm]
Li  & 2 & 1 &   -7.47806 & 0.04533 &   -7.47427(4) & 91.6 &   -7.47801(3) & 99.9 & 0.24  &       \\[.1cm]
    & 3 & 1 &            &         &   -7.47788(1) & 99.6 &   -7.47803(1) & 99.9 & 0.037 &       \\[.1cm]
    & 4 & 1 &            &         &   -7.47797(1) & 99.8 &   -7.47803(1) & 99.9 & 0.028 & 60\%\ \ \\[.1cm]
Be  & 2 & 2 &  -14.66736 & 0.09434 &  -14.66088(5) & 93.1 &  -14.66689(4) & 99.5 & 0.35  &      \\[.1cm]
    & 3 & 2 &            &         &  -14.66662(1) & 99.2 &  -14.66723(1) & 99.9 & 0.089 &      \\[.1cm]
    & 4 & 2 &            &         &  -14.66681(1) & 99.4 &  -14.66726(1) & 99.9 & 0.078 & 31\%\ \ \\[.1cm]
Ne  & 2 & 1 & -128.9376  & 0.3905  & -128.713(2)   & 42.5 & -128.919(2)   & 95.2 & 1.9\  &       \\[.1cm]
    & 3 & 1 &            &         & -128.9008(1)  & 90.6 & -128.9242(1)  & 96.6 & 0.90\ &       \\[.1cm]
    & 4 & 1 &            &         & -128.9029(3)  & 91.1 & -128.9243(8)  & 96.6 & 0.88\ &  9\%\ \ \\[.1cm]
\end{tabular}
\label{fermionic}
\end{table}

\begin{table}[tbh]
\caption[]
{Total energies of the bosonic ground states of Li, Be, and Ne
obtained in variational and diffusion Monte Carlo.  For each atom, the
first and second rows respectively contain results using wave functions
with $n$-body correlations respectively with $n=3$ and $n=4$. 
$E^{\rm VMC}$ and $E^{\rm DMC}$ are the variational and 
diffusion Monte Carlo energies. ${\bar E}^{\rm DMC}$ is the average
of the two $E^{\rm DMC}$ values which should be identical except for 
statistical errors.
$\sigma_{\rm VMC}$ is the root mean square
fluctuation of the local energy in variational Monte Carlo.
The numbers in parentheses are the statistical errors in the last digit.
The last column is the efficiency of the four-body correlations as
measured by $\eta$ in Eq.(~\ref{efficiency}) using ${\bar E}^{\rm DMC}$.
Energies are in Hartree
atomic units.}
\begin{tabular}{ccddddc}
  \multicolumn{1}{c} {\ Atom} &
  \multicolumn{1}{c} {$n$}    &
  \multicolumn{1}{c} {\hskip1.1cm $E^{\rm VMC}$}  &
  \multicolumn{1}{c} {\hskip1.1cm $E^{\rm DMC}$}  &
  \multicolumn{1}{c} {\hskip.9cm $E^{\rm VMC}$ - ${\bar E}^{\rm DMC}$} &
  \multicolumn{1}{c} {\hskip.4cm $\sigma_{\rm VMC}$} &
  \multicolumn{1}{c} {$\eta$} \\
\hline
\\[-.2cm]
Li & 3 &   -8.673920(1)  &   -8.673934(1) & 0.000013 & 0.018 &      \\[.1cm]
   & 4 &   -8.673930(1)  &   -8.673932(1) & 0.000003 & 0.011 & 77\%\ \ \\[.1cm]
Be & 3 &  -19.274357(2)  &  -19.274387(2) & 0.000030 & 0.035 &      \\[.1cm]
   & 4 &  -19.274375(2)  &  -19.274387(4) & 0.000012 & 0.021 & 60\%\ \ \\[.1cm]
Ne & 3 & -266.28411(2)   & -266.28439(2)  & 0.00030\ & 0.21\ &      \\[.1cm]
   & 4 & -266.28426(2)   & -266.28442(4)  & 0.00015\ & 0.14\ & 50\%\ \ \\[.1cm]
\end{tabular}
\label{bosonic}
\end{table}

\acknowledgments
This work is supported by the Office of Naval Research and NSF grant
DMR-9214669.  We thank Claudia Filippi for useful discussions. The
calculations were performed on the IBM SP2 computer at the Cornell
Theory Center.

\appendix

\section{Invariants: combinatorics}
\label{app.combinatorics}

We express the exponent of the Jastrow factor as a polynomial in
(scaled) inter-particle coordinates.  For an $N$-particle system there
are $d={{N(N-1)}\over2}$ inter-particle distances.  The number of monomial
terms of degree $p$ in the polynomial is $\big({d+p-1\atop p}\big)$,
which grows
asymptotically as $p^{d-1}$ for large $p$ and as $d^p = N^{2p}$ for
large $N$.  The numbers of monomials of degrees 1 through 5 are shown
in Table~\ref{monomials} for the three- to six-body correlations.
If all $N$ particles are different then each additional monomial adds
a variational coefficient but if some of the $N$ particles are
identical, terms that are equivalent because of symmetry must have the 
same coefficients. In Table~\ref{coefficients} we show the number of
distinct coefficients required to take into account correlations
between two to four electrons and a nucleus.

It is apparent that exchange symmetry results in a considerable
reduction of the number of variational coefficients, but if one simply
computes the symmetric polynomials as general polynomials with
constrained coefficients, one does not reduce the computational
effort, which still requires a number of elementary arithmetic
operations on the order of the number of coefficients of the general
polynomial.  On the other hand, following Ref.~\onlinecite{Peter_inv},
one can speed up the computation by using results of the theory of
invariants, i.e., one can rewrite symmetric polynomials as an
unconstrained polynomials in new variables, viz., symmetrized sums of
monomials forming a finite basis of invariants.

The decrease of computational effort is a consequence of the fact that
the basis invariants themselves have to be computed only once,
whereupon a number of arithmetic operations equal to the number of
coefficients of the symmetric polynomial is required to complete the
computation\cite{footnote}. If one is dealing with a system with $N$
particles one gets the full benefit of the approach by using
invariants associated with the group associated with exchange of {\em
all} identical particles. One problem of this approach is that the basis
invariants become difficult to construct, but a more fundamental
problem is that this approach automatically incorporates all
$N$-body correlations, in spite of our expectation that $n$-body
correlations become rapidly less important as $n$ increases.  We have
therefore implemented a hybrid approach which consists of using
$n$-body invariants ($n=3$ or $n=4$ for the results reported in this
paper) and symmetrizing the resulting expressions over all possible
$\big({N \atop n-1}\big)$ choices of the electrons.
A method of construction of these invariants and further details 
will be published elsewhere\cite{Speedup}.

In addition to the above symmetry considerations concerning the
reduction of the number of free variational parameters, we mention
that imposition of the cusp conditions\cite{Kato} has the additional
advantage of ensuring that the local energy is finite at particle
coincidences.  Table~\ref{free} displays the number of free
variational coefficients, after imposition of the cusp conditions,
when two electrons and a nucleus (e$^2$-n) and three electrons and a
nucleus (e$^3$-n) are correlated.  Comparison of
Tabs.~\ref{coefficients} and \ref{free} shows that a large reduction
in the number of variational coefficients is achieved.

\begin{table}[htb]
\caption
{Number of monomials of degree $p$ in $d$ inter-particle distances
correlating $n$ particles.}
\begin{tabular}{ l c  r r   r  r  r  r  r  r  r}
   &  &  &  & \multicolumn{6}{c}{Number of terms of polynomial order}\\
\hline
\\[-.2cm]
& $n$ &$d$ & &$p=1$ &  $p=2$ &  $p=3$ & $p=4$  &$p=5$  & Total & \\
\hline
\\[-.2cm]
& 3 &  3 & &   3  &      6 &     10 &    15  &    21 &    55 & \\[.1cm]
& 4 &  6 & &   6  &     21 &     56 &   126  &   252 &   461 & \\[.1cm]
& 5 & 10 & &  10  &     55 &    220 &   715  &  2002 &  3002 & \\[.1cm]
& 6 & 15 & &  15  &    120 &    680 &  3060  & 11628 & 15503 & \\[.1cm]
\end{tabular}
\label{monomials}
\end{table}

\begin{table}[htb]
\caption{
Number of symmetrized monomials of degree $p$ in $d$ inter-particle
distances correlating $n$ particles; symmetrization is with respect to
electron interchange. Spin-up and spin-down electrons are treated as being 
identical.}
\begin{tabular}{ l c  r r   r  r  r  r  r  r  r}
   &  &  &  & \multicolumn{6}{c}{Number of terms of polynomial order}\\
\hline
\\[-.2cm]
& $n$  & $d$ & & p=1  &  p=2   &    p=3 &   p=4 &   p=5 & Total & \\
\hline
\\[-.2cm]
& 3 (e$^2$-n) &  3 &  &   2  &    4   &      6 &     9 &    12 &    33 & \\[.1cm]
& 4 (e$^3$-n) &  6 &  &   2  &    6   &     14 &    28 &    54 &   104 & \\[.1cm]
& 5 (e$^4$-n) & 10 &  &   2  &    7   &     20 &    53 &   125 &   207 & \\[.1cm]
\end{tabular}
\label{coefficients}
\end{table}

\begin{table}[htb]
\caption
{Number of free parameters associated with terms of degree $p$
taking into account the reduction from imposing the cusp conditions.
Spin-up and spin-down electrons are treated as being identical.  The
asterisk for $p=1$ indicates that the number of free parameters in the
Jastrow factor is one rather than zero because the e-n cusp condition
is satisfied by fixing one of the parameters in each of the orbitals
rather than by fixing one of the parameters in the Jastrow factor.}
\begin{tabular}{ l c  r r   r  r  r  r  r  r  r}
   &  &  &  & \multicolumn{6}{c}{Number of terms of polynomial order} \\
\hline
\\[-.2cm]
& $n$   & d &  & p=1   &p=2   &  p=3 &  p=4 &  p=5 & Total & \\
\hline
\\[-.2cm]
& 3 (e$^2$-n) & 3 &  &  1$^*$&  2   &    4 &    7 &   10 &    23 &  \\[.1cm]
& 4 (e$^3$-n) & 6 &  &  1$^*$&  2   &    5 &   13 &   31 &    51 &  \\[.1cm]
\end{tabular}
\label{free}
\end{table}

\section{Basis invariants}
\label{inv}

In this appendix we discuss the computation of the polynomials in the
Jastrow factor and we list the invariants used in our computations.

The exponent of the generalized Jastrow factor is of the following
form
\begin{equation}
P(p_1, p_2, \dots p_{I})=\sum_{j_1,j_2,\cdots,j_I}
     c_{j_1,j_2,\cdots,j_I}
     p_1^{j_1} p_2^{j_2} \cdots p_I^{j_I}.
\end{equation}
which is a polynomial of the $I$ basis polynomials $p_1,p_2\dots
p_{I}$ and can be computed efficiently using a scheme derived from the
recursive, multi-variate generalization of Horner's rule \cite{Horner}.
That is, consider $P$ to be a polynomial in $p_I$ and apply Horner's rule 
for its evaluation.  The coefficients of this polynomial in $p_I$ are
polynomials in $p_1,\cdots,p_{I-1}$ which are computed recursively,
using the same scheme.

We use a variant of this algorithm designed to evaluate the monomials
in $P$ separately, rather than just $P$ itself.  The reason is that we
use a fixed set of configurations for optimizing the trial wave
functions, so that the monomials comprising the above polynomial do
not change during the optimization provided that the scale factor
$\kappa$ in Eq.~(\ref{eq.scaleR}) is kept constant.  We note
parenthetically that fixing the value of $\kappa$ does not result in a
large reduction in the variational freedom for a significant range of
values of $\kappa$ around the optimal value.  Hence, it is
computationally efficient to save the values of the monomials for each
configuration of the sample used for the wave function optimization,
and to evaluate the value of the polynomial $P$ for each configuration
as a dot product of this constant vector of monomials with the varying
coefficient vector $c_{j_1,j_2,\cdots,j_I}$.

Our variant of Horner's rule is as follows:  We start with the basis
invariants $p_1,\cdots,p_I$ and regard them as of degree one.
At the second stage, we construct all possible polynomials
quadratic in the basis invariants by multiplying all invariants of
first degree by $p_1$, then multiply all except $p_1$ by $p_2$, all
except $p_1$ and $p_2$ by $p_3$ and so on and so forth.

At stage $n$ of the calculation we construct monomials of degree $n$ in
the basis invariants $p_1,\cdots,p_I$ from those of degree $n-1$ obtained at
the previous stage by multiplying (1) all
the monomials of degree $n-1$ by $p_1$; (2) all except those descended
from $p_1$ by $p_2$; (3) all except those descended from $p_1$ and
$p_2$ by $p_3$; {\it etc.}  Hence, we obtain the following sequence of
monomials:
\begin{eqnarray}
v_1&=&(p_1; \;p_2; \;p_3; \;\cdots; \;p_I), \\
v_2&=&(p_1^2,p_1 p_2,p_1 p_3, \;\cdots;
      \;p_2^2,p_2 p_3 \;\cdots;
      \;p_3^2 \;\cdots;
      \;\cdots; \;p_I^2), \\
v_3&=&(p_1^3,p_1^2 p_2,p_1^2 p_3,\;\cdots,p_1 p_2^2,p_1 p_2 p_3,\;\cdots,p_1 p_3^2 \;\cdots;
       \;p_2^3,p_2^2 p_3,\;\cdots,p_2 p_3^2 \;\cdots;
       \;p_3^3 \;\cdots;
       \; \;\cdots; \;p_I^3), \\
&\vdots \nonumber \\
v_n&=&(p_1^n,p_1^{n-1}p_2,\cdots; \;p_2^n,p_2^{n-1}p_3,\cdots;
        \;p_3^n,\cdots; \;\cdots; \;p_I^n).
\end{eqnarray}
Generalization of the above algorithm to the computation of the gradient
and Laplacian required in the computation is straightforward.
The basis invariants as functions of the scaled inter-particle
distances are homogeneous polynomials of various degrees, and in
practical applications one truncates the polynomial $P$ at some chosen
degree in the inter-particle distances. The above algorithm, however,
will generate monomials of varying degrees in these at any step.  This
can be corrected by simply not constructing monomials that exceed the
maximal degree.

For the case of two identical electrons (labeled by $i$ and $j$) and a
nucleus (labeled by $\alpha$) the following set of basis invariants can be
used:
\begin{eqnarray}
p_1 &=& r_{ij} \\
p_2 &=& r_{\alpha i} + r_{\alpha j}, \\
p_3 &=& r_{\alpha i} r_{\alpha j}.
\end{eqnarray}

For three electrons (denoted by subscripts $i,j,k$) and a nucleus
(denoted by $\alpha$) we
used:
\begin{eqnarray}
p_1 &=& r_{ij} + r_{ik} + r_{jk} \\
p_2 &=& r_{\alpha i} + r_{\alpha j} + r_{\alpha k} \\
p_3 &=& r_{ij}^2 + r_{ik}^2 + r_{jk}^2  \\
p_4 &=& r_{\alpha i}^2 + r_{\alpha j}^2 + r_{\alpha k}^2  \\
p_5 &=& r_{\alpha k}r_{ij} + r_{\alpha j}r_{ik} + r_{\alpha i}r_{jk}  \\
p_6 &=& r_{ij}r_{ik}r_{jk}  \\
p_7 &=& r_{\alpha i}r_{\alpha j}r_{\alpha k}  \\
p_8 &=& r_{\alpha i}r_{\alpha j}r_{ij} + r_{\alpha i}r_{\alpha k}r_{ik} +
        r_{\alpha j}r_{\alpha k}r_{jk}  \\
p_9 &=& r_{\alpha i}r_{ij}r_{ik} + r_{\alpha j}r_{ij}r_{jk} +
        r_{\alpha k}r_{ik}r_{jk}
\end{eqnarray}

Finally, we note that the above choice of basis invariants is not
unique.  They were constructed to yield invariants consisting of a small
number of monomials each\cite{Speedup}.  It should be noted that the
we have no proof that the nine invariants given above indeed form a
complete basis in the mathematical sense, but they are complete as far
our current computations are concerned.


\begin{references}
\bibitem{footnote} Strictly speaking this statement is correct only if the
basis invariants are algebraically independent.  In practice, the
algebraic dependences introduce only very few additional coefficients
in the range of polynomial degrees of practical interest.
\end{references}
\end{document}